\documentstyle[prb,aps,epsfig]{revtex}
\begin{document}
\draft
\title{Bond breaking in vibrationally excited methane on transition
  metal catalysts}
\author{R. Milot and A.~P.~J. Jansen}
\address{Schuit Institute of Catalysis, ST/SKA, Eindhoven University of
  Technology\\ P.O. Box 513, NL-5600 MB Eindhoven, The Netherlands}
\date{\today}
\maketitle

\begin{abstract}
  The role of vibrational excitation of a single mode in the scattering
  of methane is studied by wave packet simulations of oriented CH${}_4$
  and CD${}_4$ molecules from a flat surface. All nine internal
  vibrations are included. In the translational energy range from 32 up
  to 128 kJ/mol we find that initial vibrational excitations
  enhance the transfer of translational energy towards vibrational
  energy and increase the accessibility of the entrance channel for
  dissociation.  Our simulations predict that initial vibrational
  excitations of the asymmetrical stretch ($\nu_3$) and especially the
  symmetrical stretch ($\nu_1$) modes will give the highest enhancement
  of the dissociation probability of methane.
\end{abstract}

The dissociative adsorption of methane on transition metals is an
important reaction in catalysis; it is the rate limiting step in steam
reforming to produce syngas, and it is prototypical for catalytic C--H
activation. Although the reaction mechanism has been studied
intensively, it is not been fully understood yet. A number of molecular
beam experiments in which the dissociation energy was measured as a
function of translational energy have observed that vibrationally hot
CH${}_4$ dissociates more readily than cold CH${}_4$, with the energy in
the internal vibrations being about as effective as the translational
energy in inducing
dissociation.\cite{ret85,ret86,lee87,lun89,hol95,lar99,walker99} Two
independent bulb gas experiment with laser excitation of the $\nu_3$
asymmetrical stretch and $2\nu_4$ umbrella modes on the Rh(111)
surface,\cite{yates79} and laser excitation of the $\nu_3$ and $2\nu_3$
modes on thin films of rhodium \cite{brass79} did not reveal any
noticeable enhancement in the reactivity of CH${}_4$.  A recent
molecular beam experiment with laser excitation of the $\nu_3$ mode did
succeed in measuring a strong enhancement of the dissociation on a
Ni(100) surface. However, this enhancement was still much too low to
account for the vibrational activation observed in previous studies and
indicated that other vibrationally excited modes contribute
significantly to the reactivity of thermal samples.\cite{juur99}

Wave packet simulations of the methane dissociation reaction on
transition metals have treated the methane molecule always as a diatomic
up to now.\cite{har91,lun91,lun92,lun95,jan95,car98} Apart from one
C--H bond (a pseudo $\nu_3$ stretch mode) and the molecule surface
distance, either (multiple) rotations or some lattice motion were
included. None of them have looked at the role of the other internal
vibrations, so there is no model that describes which vibrationally
excited mode might be responsible for the experimental observed
vibrational activation. In previous papers we have reported on wave
packet simulations to determine which and to what extent internal
vibrations are important for the dissociation of CH${}_4$ in the
vibrational ground state,\cite{mil98} and the isotope effect of
CD${}_4$.\cite{mil00a} We were not able yet to simulate
the dissociation including all internal vibrations. Instead we simulated
the scattering of methane, for which all internal vibrations can be
included, and used the results to deduce consequences for the
dissociation. These simulations indicate that for methane to dissociate
the interaction of the molecule with the surface should lead to an
elongated equilibrium C--H bond length close to the surface. In this
letter we report on new wave packet simulations of the role of
vibrational excitations for the scattering of CH${}_4$ and CD${}_4$
molecules with all nine internal vibrations. The dynamical features of
these simulations give new insight into the initial steps of the
dissociation process. The conventional explanation is that vibrations
help dissociation by adding energy needed to overcome the dissociation
barrier.  We find that two other new explanations play also a role. One
of them is the enhanced transfer of translational energy into the
dissociation channel by initial vibrational excitations. The other more
important explanation is the increased accessibility of the entrance
channel for dissociation.

We have used the multi-confi\-guratio\-nal time-depen\-dent Hartree (MCTDH)
method for our wave packet simulation.\cite{man92,jan93} This method
can deal with a large number of degrees of freedom and with large grids.
(See Ref.~\onlinecite{bec00} for a recent review.)  Initial
translational energy has been chosen in the range of 32 to 128 kJ/mol.
The initial state has been written as a product state of ten functions;
one for the normally incident translational coordinate, and one for each
internal vibration. All vibrations were taken to be in the ground state
except one which was put in the first excited state. The orientation of
the CH${}_4$/CD${}_4$ was fixed, and the vibrationally excited state had
$a_1$ symmetry in the symmetry group of the molecule plus surface
(C${}_{3v}$ when one or three H/D atoms point towards the surface, and
C${}_{2v}$ when two point towards the surface.) The potential-energy
surface is characterised by an elongation of the C--H bonds when the
molecule approaches the surfaces, no surface corrugation, and a
molecule-surface part appropriate for Ni(111). It has been shown to give
reasonable results, and is described in Refs.~\onlinecite{mil98} and
\onlinecite{mil00a}. These articles give also the computational details
about the configurational basis and number of grid points, and contain
illustrations of the orientations and the important vibrational modes.

\begin{figure}[h]
  \epsfig{file=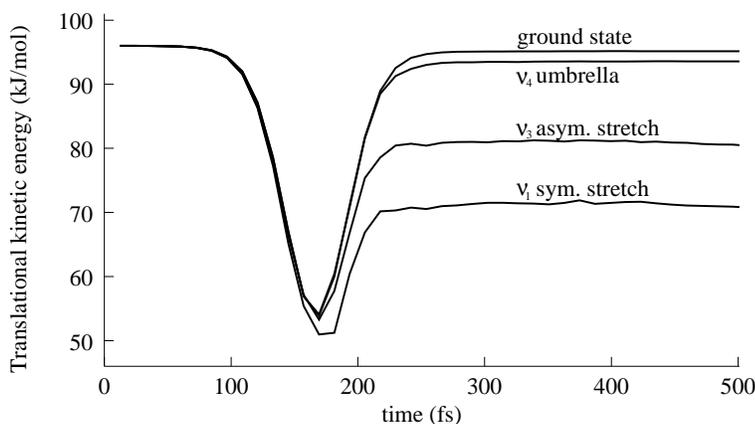, height=5.5cm}
  \caption{Translational kinetic energy versus time for a CH${}_4$
    molecule with three bonds pointing towards the surface. The initial
    translational kinetic energy is 96 kJ/mol.}
  \label{fig:trans96}
\end{figure}

We can obtain a good idea about the overall activation of a mode by
looking at the kinetic energy expectation values $\langle\Psi (t)\vert
T_j \vert\Psi (t)\rangle$ for each mode $j$. During the scattering
process the change in the translational kinetic energy is the largest.
It is plotted in Fig.~\ref{fig:trans96} as a function of time for
CH${}_4$ in the orientation with three bond pointing towards the surface
with an initial kinetic energy of 96 kJ/mol and different initial
vibrational states. When the molecule approaches the surface the kinetic
energy falls down to a minimum value.  This minimum value varies only
slightly with the initial vibrational states of the molecule. The total
loss of translational kinetic energy varies substantially, however. The
initial translational kinetic energy is not conserved. This means that
the vibrational excitation enhances inelastic scattering.  Especially an
excitation of the $\nu_1$ symmetrical stretch and to a lesser extend the
$\nu_3$ asymmetrical stretch mode result in an increased transfer of
kinetic energy towards the intramolecular vibrational energy. The
inelastic scatter component (the initial minus the final translational
energy) for both isotopes in the orientation with three bonds pointing
towards the surface, shows the following trend for the initial
vibrational excitations of the modes; $\nu_1$ $>$ $\nu_3$ $>$ $\nu_4$
$>$ ground state. CH${}_4$ scatters more inelastic than CD${}_4$ over
the whole calculated range of translational kinetic energies, if the
molecule has an initial excitation of the $\nu_3$ stretch mode.
CH${}_4$ scatters also more inelastically than CD${}_4$ in the $\nu_1$
symmetrical stretch mode at higher energies , but at lower energies it
scatters slightly less inelastically.  For the molecules with the
non-excited state or an excitation in the $\nu_4$ umbrella mode CD${}_4$
has a higher inelastic scattering component than CH${}_4$. At an initial
translational kinetic energy of 128 kJ/mol the excitation of the $\nu_4$
umbrella mode results in a strong enhancement of the inelastic
scattering component. For CD${}_4$ the inelastic scattering component
for the initial excited $\nu_4$ umbrella mode can become even larger
than for the initial excited $\nu_3$ stretch mode.  For the orientation
with two bonds pointing towards the surface we observe the same trends
for the relation between the inelastic scatter components and the
excited initial vibrational modes, but the inelastic scatter component
are less than half of the values for the orientation with three bonds
pointing towards the surface. Also the excitation of the $\nu_3$
asymmetrical stretch modes results now in a higher inelastic scattering
component for CD${}_4$ than for CH${}_4$. Excitation of the $\nu_2$
bending mode gives a little higher inelastic scatter component than the
vibrational ground state. For the orientation with one bond pointing
towards the surface we observe an even lower inelastic scattering
component. At an initial kinetic energy of 128 kJ/mol we find that both
the $\nu_1$ and $\nu_3$ stretch modes have on inelastic component of
around 6.5 kJ/mol for CD${}_4$ and 4.0 kJ/mol for CH${}_4$. At an
initial translational energy of 32 kJ/mol we observe for both isotopes
in all orientations a very small increase of translational kinetic
energy for the vibrational excited molecule, which means that there is a
net transfer from intramolecular vibrational energy through the surface
repulsion into the translational coordinate.

There seem to be two groups of vibrations with different qualitative
behavior with respect to (de)excitation when the molecule hits the
surface. The first group, let's call it the ``stretch'' group, consist
of the $\nu_3$ asymmetric stretch in any orientation and the $\nu_1$
symmetric stretch in the orientation with three hydrogen/deuterium atoms
pointing to the surface. The second, let's call it the ``bending''
group, consists of all bending vibrations and the $\nu_1$ in other
orientations. When the molecule is initially in the vibrational ground
state the kinetic energy in the vibrations increases, reaches a maximum
at the turn-around point, and then drops back almost to the initial
level except for a small contribution due to the inelastic scattering
component. The vibrations within a group have very similar amounts of
kinetic energy, but the ``stretch'' group has clearly a larger inelastic
component than the ``bending'' group, and also the kinetic energy at the
turn-around point is larger. When the molecule has initially an
excitation of a vibration of the ``stretch'' group then the kinetic
energy of that vibration increases, reaches a maximum at the turn-around
point, and drops to a level lower than it was initially. For an
excitation of a vibration of the ``bending'' group there is no maximum,
but its kinetic energy simply drops to a lower level. We see that in all
cases there is not only a transfer of energy from the translation to
vibrations, but also an energy flow from the initially excited vibration
to other vibrations. However, the total energy of the vibrational
kinetic energy and the intramolecular potential energy
increases, because it has to absorb the inelastic scattering component.

\begin{figure}[h]
  \epsfig{file=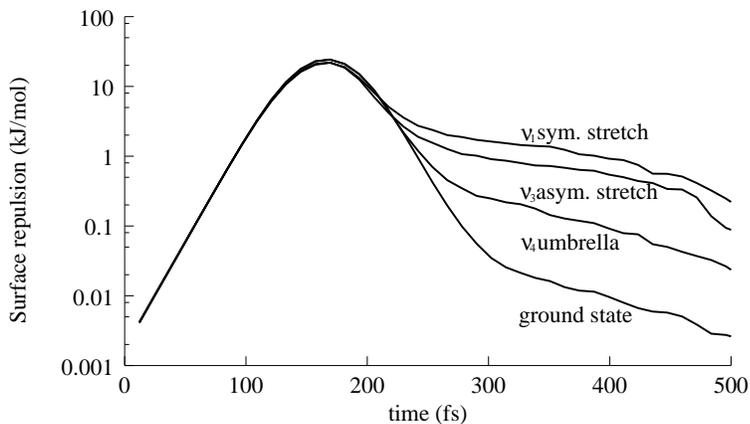, height=5.5cm}
  \caption{Surface repulsion versus time during the scattering dynamics
    of CH${}_4$ at an initial translational energy of 96 kJ/mol in the
    orientation with three bonds pointing towards the surface. }
  \label{fig:vsurf}
\end{figure}

Figure \ref{fig:vsurf} shows the (repulsive) interaction with the
surface during the scattering process of CH${}_4$ at an initial kinetic
energy of 96 kJ/mol and different initial vibrational excitations for
the orientation with three hydrogens pointing towards the surface. Since
this is a repulsive term with a exponential fall-off changes in the
repulsion indicate the motion of the part of the wave packet closest to
the surface.  At the beginning of the simulation the curves are almost
linear in a logarithmic plot, because the repulsion hardly changes the
velocity of the molecule.  After some time the molecule enters into a
region with a higher surface repulsion and the slopes of the curves
drop. This results in a maximum at the turn-around point, where most of
the initial translational kinetic energy is transfered into potential
energy of the surface repulsion.  For a classical simulation it would
have meant no translational kinetic energy, but it corresponds with the
minimum kinetic energy for our wave packet simulations.  Past the
maximum, a part of the wave packet will accelerate away from the
surface, and the slope becomes negative.  The expectation value of the
translational kinetic energy (see Fig.\ref{fig:trans96}) increases at
the same time.  The slope of the curves in Fig.~\ref{fig:vsurf} becomes
less negative towards the end of the simulation, although the
expectation value of the translational kinetic energy in this time
region is almost constant.  The reason for this is that a part of the
wave packet with less translational kinetic energy is still in a region
close to the surface. We see also that the height of the plateaus for
the different initial vibrational excitations is again in the order;
$\nu_1$ $>$ $\nu_3$ $>$ $\nu_4$ $>$ ground state. This again indicates
that a larger part of the wave packet is inelastically scattered when
$\nu_1$ is excited than when $\nu_3$ is excited, etc.

At lower initial translational kinetic energies the plateaus have a
lower position and the main gap exist between the plateaus of the
$\nu_1$ and $\nu_3$ stretch modes and the lower positioned plateaus of
the $\nu_4$ umbrella and the ground state. At an initial translational
kinetic energies of 128 kJ/mol the positions of the plateaus are higher
and the differences between the initial vibrational excitations are also
smaller.  The plateau of the $\nu_3$ stretch mode is even around the
same position as the $\nu_4$ umbrella mode for CD${}_4$ in the
orientation with three bonds pointing towards the surface at this
initial energy.  The orientation with two bonds pointing towards the
surface shows the same trends. The plateaus of the initial excited
$\nu_2$ bending mode are located slightly above the ground state for
both isotopes.  For the orientation with one bond pointing towards the
surface the relative positions of the plateaus of the different initial
excitations are the same as at low energies in the orientation with
three bonds pointing towards the surface.
 
Even though we did not try to describe the dissociation itself, the
scattering simulation do yield indications for the role of vibrational
excitations on the dissociation of methane, and compare these with
experimental observations.  The dissociation of methane occurs over a
late barrier, because it is enhanced by vibrational energy.\cite{lev87}
Conventionally, the role of vibrational excitation on the enhancement of
dissociation probability was discussed as an effect of the availability
of the extra (vibrational kinetic) energy for overcoming the
dissociation barrier. Our simulations show that such a process might
play a role, but they show also that two other processes occur through
vibrational excitation.

Firstly, an initial vibrational excitation increases translational
kinetic energy transfer towards the intramolecular vibrational energy.
The simulations show that this inelastic scatter component can be seen
in an large enhancement of the vibrational kinetic energy in the stretch
modes at the turn-around point.  This increase is larger for higher
initial translational kinetic energy and is most effective in the
orientation with three bonds pointing towards the surface. If the
dissociation of methane occurs primarily in this orientation, then we
would expect, based on the total available vibrational energy after
hitting the surface, that excitation of the $\nu_1$ symmetrical stretch
mode is the most effective for enhancing the dissociation probability.
The $\nu_3$ asymmetrical stretch mode appears to be less so. An
explanation of the enhanced inelastic scatter compound by vibrational
excitation is that through excitation the bonds are weakened, which will
ease excitation in the initial non-excited modes. Other excitations than
the $\nu_2$, $\nu_3$, or $\nu_4$ with $a_1$ symmetry for a particular
orientation can possibly result in higher energy transfers, but we think
that the difference with $\nu_1$ (which has always $a_1$ symmetry) would
be still large.

Secondly, the accessibility of the dissociation channel enhances also
the dissociation probability. We have concluded previously that our
potential mimics reasonably the entrance channel for
dissociation.\cite{mil98} In this letter we find that a part of the wave
packet has a longer residence time at the surface. It is this part of
the wave packet that accesses the dissociation channel, and it is also this
part that is able to come near to the transition state for dissociation.
From Figs.~\ref{fig:trans96} and \ref{fig:vsurf} we conclude that the
$\nu_1$ stretch mode will enhance the dissociation probability the most.
The enhanced accessibility by vibrational excitation is explained by the
spread of the wave packet along a C--H bond, which gives a higher
probability for the system to be atop the dissociation barrier.

The molecular beam experiment with excitations of the $\nu_3$
asymmetrical stretch mode of CH${}_4$ of Ref.~\onlinecite{juur99} shows
that a single excitation of the $\nu_3$ asymmetrical stretch mode
enhances dissociation, but the measured reactivity of the $\nu_3$
stretch mode is too low to account for the total vibrational activation
observed in the molecular beam study of Ref.~\onlinecite{hol95}. It
means that excitation of another mode than the $\nu_3$ stretch will be
more effective for dissociation. Our simulations show that indeed
excitation of $\nu_3$ stretch will enhance dissociation, but predict
that excitation of the $\nu_1$ symmetrical stretch mode will be more
effective if the dissociation occurs primary in the orientation with
multiple bonds pointing towards the surface. The contribution of the
$\nu_1$ symmetrical stretch mode cannot be measured directly, because it
has no infra-red activity. However, the contribution of the $\nu_1$ mode
can be estimated using a molecular beam study as follows. The
contribution of the $\nu_3$ stretch has already been determined.
\cite{juur99} Similarly the contribution of the $\nu_4$ umbrella mode
can be determined. The contribution of the $\nu_2$ bending can be
estimated from our simulations to be somewhat lower than the $\nu_4$
umbrella contribution. The total contribution of all vibrations is known
from Ref.~\onlinecite{hol95}, and a simple subtraction will give us then
the contribution of the $\nu_1$ stretch.  At high translational energies
the accessibility of the dissociation channel for molecules with an
excited $\nu_4$ umbrella mode is near to that of the molecules with
excited stretch modes, and for CD${}_4$ the inelastic scattering is also
enhanced.  So the excitation of the $\nu_4$ umbrella mode can still
contribute significantly to the vibrational activation, because it has
also higher Boltzmann population in the molecular beam than the stretch
modes.

This research has been financially supported by the Council for Chemical
Sciences of the Netherlands Organization for Scientific Research
(CW-NWO), and has been performed under the auspices of the Netherlands
Institute for Catalysis Research (NIOK).

\end{document}